\documentclass[12pt]{article}
\usepackage{amsmath}
\usepackage{epsfig}
\usepackage{amsfonts}
\usepackage{feynmp}
\usepackage{latexsym}

 \title{On threshold amplitudes III: $2\rightarrow n$\ processes} 
\author{Joanna Domienik\thanks{supported by the {\L}\'od\'z University grant N$^0$\ 269 },\\ 
 Piotr Kosi\'nski\thanks{supported by KBN grant N$^0$\ 5 P03B 060 21 } \\
Department of Theoretical Physics II \\
University of {\L}\'od\'z \\
Pomorska 149/153, 90 - 236 {\L}\'od\'z/Poland.}
 \date{} 
\begin{document} 
\maketitle  
\begin{abstract} 
The $2\rightarrow n$\ scattering with final particles at rest is discussed. The comparison with purely soft processes allows
to identify symmetries responsible for vanishing of certain $2\rightarrow n$\ amplitudes. Some examples are given.
\end{abstract} 
\newpage
\section{Introduction}
$ $\

In the present paper, third of the series (cf. \cite{b1}, \cite{b2}), we consider $2\rightarrow n$\ processes: two 
particles of opposite momenta scatter to produce $n$\ bosons at rest. This process has been already considered, on 
the tree level, using propagator approach \cite{b3} $\div$\ \cite{b6}, diagrammatics \cite{b7}, \cite{b8} or Feynman
wave function method \cite{b9}, \cite{b1}. The last method seems to be the most efficient. 

In Sec. II we present 
a very simple and straightforward derivation of the Brown-Zhai \cite{b9} algorithm. We give a general procedure to construct
$2\rightarrow n$\ amplitudes which, in particular, allows us, for a wide class of theories, to compare $2\rightarrow n$\ 
amplitude with a purely soft one. By using the results concerning the relations between symmetries and vanishing of
purely soft amplitudes \cite{b10}, \cite{b2} we are able to identify, in certain cases, the symmetries underlying
nullification in $2\rightarrow n$\ scattering. 

In Sec. III we discuss some examples involving only bosons while Sec. IV is devoted to fermion-antifermion scattering. The
latter is discussed in more detail and the complete expression for the amplitude is given and compared, for $n=2$,
with diagrammatic result. 

Sec. V contains some conclusions.
\newpage
\section{Calculating $2\rightarrow n$\ amplitudes}
$ $\

Let us first remind the general procedure for calculating the matrix elements of field operators in the tree approximation
\cite{b9}, \cite{b1}. Our starting point is a lagrangian
\begin{eqnarray}
L=\frac{1}{2}\sum\limits_{i=1}^N(\partial_{\mu}\Phi_i\partial^{\mu}\Phi_i-m^2_i\Phi_i^2)-V(\underline{\Phi}),\label{w1}
\end{eqnarray}
where $\{\Phi_i\}^N_{i=1}$\ is a multiplet of scalar fields; higher-spin fields can be also included.

Let $\{f_n^{(i)}\}_{n=0}^{\infty}$\ be a complete set of normalized positive-energy solutions to the Klein-Gordon equation
of mass $m_i$; in what follows we shall use momentum basis, $f_{\vec{p}}^{(i)}(x)=\frac{1}{\sqrt{(2\pi  )^32E_p}}e^{-ipx}$,
$p^2=m_i^2$. Define the classical free field by
\begin{eqnarray}
\Phi_{0i}(x)=\sum_n(\beta_n^{(i)}f_n^{(i)}(x)+\bar{\beta}_n^{(i)}\overline{f_n^{(i)}(x)}) \label{w2}
\end{eqnarray}
Consider the set of integral equations for $\Phi_i(x)$,
\begin{eqnarray}
\Phi_i(x)=\Phi_{0i}(x)-\int d^4y\Delta_{F_{ij}}(x-y)\frac{\partial V}{\partial \Phi_j(y)}\label{w3}
\end{eqnarray}
where $\Delta_{F_{ij}}(x)=\Delta_F(x;\;m_i^2)\delta_{ij}$\ is the Feynman propagator. Under some assumptions eqs. 
(\ref{w3}) admit unique solution, at least at the formal perturbative level. Once eq. (\ref{w3}) is solved we 
obtain $\Phi_i(x)$\ as a function of $\beta_n^{(i)}$\ and $\bar{\beta}_n^{(i)}$,
\begin{eqnarray}
\Phi_i(x)\equiv \Phi_i(x\mid \beta ,\;\bar{\beta})\label{w4}
\end{eqnarray}
Taking succesive derivatives with respect to $\beta$'s and $\bar{\beta}$'s at $\beta =\bar{\beta}=0$\ one obtains the matrix
elements of the field operator $\hat{\Phi}_i(x)$\ between in- and out states in the tree-level approximation \cite{b9}, 
\cite{b1}; for example,
\begin{eqnarray}
\frac{\partial^2\Phi_i(x\mid \beta ,\bar{\beta})}{\partial \beta_n^{(k)}\partial \bar{\beta}_m^{(l)}}
\hbox{\Huge $\mid$}_{\beta =\bar{\beta}=0}=<(ml);out\mid \hat{\Phi}_i(x)\mid (nk);in > \hbox{\Huge $\mid$}_{tree}\label{w5}
\end{eqnarray}

In practice, it is more convenient to work with differential equations instead of integral ones. Eqs. (\ref{w2},\ref{w3}) imply
\begin{eqnarray}
&&(\Box +m_i^2)\Phi_i(x\mid \beta , \bar{\beta})+\frac{\partial V(\Phi)}{\partial \Phi_i(x\mid \beta ,\bar{\beta})}=0\label{w6}\\
&&\Phi_i(x\mid \beta,\bar{\beta})\mid_{V=0}=\Phi_{0i}(x) \label{w7}
\end{eqnarray}

The converse is, in general, not true. Indeed, for example $\beta$'s and $\bar{\beta}$'s enter $\Phi_i(x\mid \beta,\bar{\beta})$\
as arbitrary constants; they can be replaced, without alterning (\ref{w6}) and (\ref{w7}), by any coupling constant
dependent functions $\beta (\lambda ),\; \bar{\beta}(\lambda )$\ such that $\beta (0)=\beta ,\;\bar{\beta}(0)=\bar{\beta}$. The 
solution to (\ref{w6}), (\ref{w7}) can be made unique by adding further constraints following from the tree-graph 
interpretation. Consider $\Phi^4$-theory as an example,
\begin{eqnarray}
V(\Phi )=\frac{\lambda}{4!}\Phi^4 \nonumber
\end{eqnarray}
Topological relations for tree graphs imply that $\beta,\; \bar{\beta}$\ and $\lambda$\ enter the matrix elements
of field operator in the combination 
$\lambda^n\beta^k\bar{\beta}^l$\ with $k+l=2n+1$; in other words, the contribution of the $\lambda^n$-order is a homogeneous
polynomial in $\beta$\ and $\bar{\beta}$\ of degree $2n+1$. This additional condition makes (\ref{w6}), (\ref{w7}) 
equivalent to (\ref{w3}).

Let us now apply the above formalism to the problem of $2\rightarrow n$\ scattering. Assume we have two initial particles of 
the same kind, carried, say, by $\Phi_1$, with the momenta $\vec{p}$\ and $-\vec{p}$; we want to compute the amplitude
for producing $n$ particles at rest, the final particles being of different kind than the initial ones.

Differentiating (\ref{w6}), (\ref{w7}) with respect to $\beta_{\vec{p}}^{(1)}$\ we arrive at the following set of equations
\begin{eqnarray}
&&(\Box +m_i^2)\frac{\partial \Phi_i(x)}{\partial \beta_{\vec{p}}^{(1)}}+\frac{\partial^2V}{\partial \Phi_i(x)\partial \Phi_j(x)}
\frac{\partial \Phi_j(x)}{\partial \beta_{\vec{p}}^{(1)}}=0\label{w8} \\
&&\frac{\partial \Phi_i(x)}{\partial \beta_{\vec{p}}^{(1)}}\hbox{\Huge $\mid$}_{V=0}\equiv 
\delta_{i1}f_{\vec{p}}^{(1)}(x)\label{w9}
\end{eqnarray}
Now, $\frac{\partial \Phi_j(x)}{\partial \beta_{\vec{p}}^{(1)}}$\ generates, in the tree approximation, the matrix elements
of $i$-th field operator with one "1" particle carrying the momentum $\vec{p}$\ already present in the initial state. The
second initial particle is obtained by applying LSZ reduction procedure to the matrix elements of $ \hat \Phi _1(x)$. 
Therefore, we have only to generate final  particles at the threshold and the only parameters we need to keep nonvanishing 
are $ {\overline \beta }^{(i)}_0, \;\; i=2,....,N $. This allows us to simplify the whole algorithm. First of all note 
that all matrix elements generated by the solutions to (\ref{w6}) are now $ \vec x$\ -independent and eqs. (\ref{w6}), 
(\ref{w7}) can be rewritten as
\begin{eqnarray}
(\partial _t^2+m_i^2)\Phi _i(t\mid \beta ,{\overline \beta })+\frac{\partial V(\Phi )}{\partial \Phi _i(t\mid \beta ,
{\overline \beta })}=0 \label{w10} \\
\Phi _i(t\mid \beta ,{\overline \beta })\mid _{V=0}=\frac{{\overline \beta }^{(i)}_0}{\sqrt {(2\pi )^32m_i}}e^{im_it}
\label{w11}
\end{eqnarray} 
Denote
\begin{eqnarray}
\psi _i(x\mid \beta ,{\overline \beta })\equiv \frac{\partial \Phi _i(x\mid \beta ,{\overline \beta })}{\partial \beta 
^{(1)}_{\vec p}} \; , \label{w12}
\end{eqnarray}
again with all $ \beta 's$\ and $ {\overline \beta }'s$,  except $ {\overline \beta }^{(i)}_0$, $ i\neq 1$, vanishing: 
$ \psi _i(x\mid \beta ,{\overline \beta })$\ generate matrix elements with one particle, carrying the momentum $ \vec p$, 
in the initial state and arbitrany number of particles at rest in the final one. Therefore, due to the translational 
invariance, 
\begin{eqnarray}
\psi _i(x\mid \beta ,{\overline \beta })=\tilde \psi _i(t\mid \beta ,{\overline \beta })e^{i\vec p\vec x}\label{w13}
\end{eqnarray}
and eqs. (\ref{w8}), (\ref{w9}) take form
\begin{eqnarray}
(\partial _t^2+M_i^2)\tilde \psi _i(t\mid \beta ,{\overline \beta })+\frac{\partial ^2V}{\partial \Phi _i(t)\partial \Phi 
_j(t)}\tilde \psi _j(t\mid \beta ,{\overline \beta })=0 \label{w14} \\
\tilde \psi _i(t)\mid _{V=0}=\delta _{i1}\frac{1}{\sqrt{(2\pi )^32M_1}}e^{-iM_1t} \; , \label{w15}
\end{eqnarray}
where $ M_i^2\equiv m_i^2+\vec p^2$\ 

Having solved (\ref{w14}), (\ref{w15}) one can compute the relevant amplitude from LSZ formalism:
\begin{eqnarray}
\frac{i(2\pi )^3\delta ^{(3)}(\vec p+\vec q)}{\sqrt{(2\pi )^32M_1}}\int\limits_{-\infty}^{\infty}dte^{-iM_1t}
(\partial _t^2+M_1^2)\tilde \psi _1(t\mid \beta ,{\overline \beta })\label{w16}
\end{eqnarray}
generates all $2\rightarrow n$\ amplitudes. \\
Eqs. $ (\ref{w13}) \div  (\ref{w16})$\ provide the general solution to our problem.

Particulary interesting situation emerges if the theory exhibits additional symmetry
\begin{eqnarray}
\Phi _1\longrightarrow \Phi _1'=-\Phi _1, \;\;\; \Phi _i\longrightarrow \Phi '_i=\Phi _i, \;\;\; i\neq 1 \label{w17}
\end{eqnarray}
Consider first eqs. (\ref{w6}), (\ref{w7}); once we put $ \beta _{\vec q}^{(1)}=0={\overline \beta }_{\vec q}^{(1)}$\ 
both eq. (\ref{w6}) and the initial condition (\ref{w7}) are invariant under (\ref{w17}). By uniqueness of the solution we 
find
\begin{eqnarray}
\Phi _1(x\mid \beta ,{\overline \beta })\mid _{\beta _{\vec q}^{(1)}=0={\overline \beta }_{\vec q}^{(1)}}=0\label{w18}
\end{eqnarray}
This conclusion is obviously equivalent to the statement that only graphs with even number of external "1"-lines are 
allowed. The counterparts of (\ref{w10}), (\ref{w11}) read now
\begin{eqnarray}
&&(\partial ^2_t+m_i^2)\Phi _i(t)+\frac{\partial V_1(\Phi )}{\partial \Phi _1(t)}=0\:, i=Z,....,N \label{w19} \\
&&\Phi _i(t)\mid _{V=0}={\overline \beta }^{(i)}_0e^{im_it} \; ,i=2,\;......,N \label{w20} \\
&&V_1({\underline \Phi })=V({\underline \Phi })\mid _{\Phi _1=0} \; , \label{w21}
\end{eqnarray}
which is a set of coupled dynamical equations for the reduced system obtained by ignoring $ \Phi _1$. Let us further 
consider eqs.(\ref{w14}). We show that
\begin{eqnarray}
\frac{\partial \Phi _j(x\mid \beta , {\overline \beta })}{\partial \beta _{\vec p}^{(1)}}
\hbox{\Huge $\mid$}_{\beta _{\vec q}^{(1)}=0={\overline \beta }_{\vec q}^{(1)}}=0 \;\; ,\; j\neq 1 \label{w22}
\end{eqnarray}
This is, again, a consequence of the fact that only even number of external "1"-lines is admitted. The formal proof 
is based on symmetry (\ref{w17}): from the uniqueness  of the solution to (\ref{w6}), (\ref{w7}) it follows then that 
$ \Phi _i(x\mid \beta ,{\overline \beta }), \;\; i\neq 1$, are even functions of $ \beta _{\vec q}^{(1)}, \;\; {\overline 
\beta }_{\vec q}^{(1)}$\ which implies (\ref{w22}).

Taking into account the above property we can rewrite (\ref{w14})
\begin{eqnarray}
&&(\partial _t^2+M_1^2)\tilde \psi _1(t\mid \beta ,{\overline \beta })+\frac{\partial ^2V}{\partial \Phi _1(t)^2}
\hbox{\Huge $\mid$}_{\Phi _1=0} \cdot {\tilde \psi _1(t\mid \beta ,{\overline \beta })}=0 \label{w23} \\
&&\tilde \psi _1(t)\mid _{V=0}=\frac{1}{\sqrt{(2\pi )^32M_1}}e^{-iM_1t} \label{w24}
\end{eqnarray}
We see that in the presence of the symmetry (\ref{w17}) the problem gets simplifield: first we solve N-1 equations 
for $ \Phi _i(t\mid \beta ,{\overline \beta }), \;\; i=2,.....,N$\ and then one equation (\ref{w23}) for $ \tilde \psi _1
(t\mid \beta ,{\overline \beta })$.

Let us now make the following important observation. Take the modified theory given by the lagrangian
\begin{eqnarray}
\tilde L=\frac{1}{2}(\partial _{\mu} \Phi _1\partial ^{\mu }\Phi _1-M_1^2\Phi _1^2)+\frac{1}{2}\sum\limits_{i=2}^N
(\partial _{\mu }\Phi _i\partial ^{\mu }\Phi _i-m_i^2\Phi_i^2)-V({\underline \Phi }), \label{w25}
\end{eqnarray}
with $M_1^2\equiv m_1^2+\vec p^2$, and consider the same process $ 2\rightarrow  n$\ except that now {\underline all} 
particles, both initial as well as final, are at rest. Repeating the procedure outlined above we conclude that the 
amplitudes for both processes coincide, up to irrelevant delta function expressing momentum conservation. This result 
can again be easily understood on the diagrammar level. Symmetry (\ref{w17}) implies that there is a unique chain of 
$ \Phi _1$-lines  connecting both initial particles and the three-momentum flows only through lines of this chain.

The observation made above allows us to explain, at least in some cases, the origin of amplitudes nullification in 
$ 2\rightarrow n$\ processes. In fact, for purely threshold amplitudes their nullification may result from symmetry 
properties of the reduced hamiltonian system obtained by ignoring space-dependence in the original problem \cite {b9}.  

\section {Some examples}
\subsection {Quartic interaction : 0(2)-theory}
Let us start with the model considered in \cite {b9}:
\begin{eqnarray}
L=\frac{1}{2}\sum\limits_{i=1}^{2}(\partial _\mu \Phi _i\partial ^\mu \Phi _i-m_i^2\Phi _i^2)-\frac{\lambda }{4!}
(\Phi _1^2+\Phi _2^2) \label{w26}
\end{eqnarray}
The reduced theory is integrable and separable in elliptic coordinates \cite {b11}; in fact, it is an example of the 
so-called Garnier system \cite {b11}. In order to calculate amplitudes $ "1"\rightarrow "2"$\ at the threshold we look 
for the solution to the field equations  obeying
\begin{eqnarray}
&&\Phi _1(t)\mid _{\lambda =0}=\frac{\beta _1e^{-im_1t}}{\sqrt{(2\pi )^32m_1}}\equiv z_1(t) \label{w27} \\
&&\Phi _2(t)\mid _{\lambda =0}=\frac{{\overline \beta }_2e^{im_2t}}{\sqrt{(2\pi )^32m_2}}\equiv z_2(t) \label{w28}
\end{eqnarray}
They read \cite {b12}
\begin{eqnarray}
&& \Phi _1=z_1(1-\frac{\lambda \kappa }{48m_2^2}z_2^2)(1-\frac{\lambda }{48m_1^2}z_1^2-\frac{\lambda }{48m_2^2}z_2^2+
\frac{\lambda ^2\kappa ^2}{48^2m_1^2m_2^2}z_1^2z_2^2)^{-1} \label{w29} \\
&& \Phi _2=z_2(1+\frac{\lambda \kappa }{48m_1^2}z_1^2)(1-\frac{\lambda }{48m_1^2}z_1^2-\frac{\lambda }{48m_2^2}z_2^2+
\frac{\lambda ^2\kappa ^2}{48^2m_1^2m_2^2}z_1^2z_2^2)^{-1}\; , \label{w30}
\end{eqnarray}
where $ \kappa \equiv \frac{m_1+m_2}{m_1-m_2}$.

In order to calculate $ 2\rightarrow n$\ amplitude, with all particles at rest, we compute
\begin{eqnarray}
\frac{\partial \Phi _1}{\partial \beta _1}\mid _{\beta _1=0}=(1-\frac{\lambda \kappa z_2^2}{48m_2^2})(1-\frac{\lambda }
{48m_2^2}z_2^2)^{-1}\frac{e^{-im_1t}}{\sqrt{(2\pi )^32m_1}} \label {w31}
\end{eqnarray}
On the other hand consider $ 2\rightarrow n$\ process with hard initial particles. The relevant amplitude can be computed 
according to the scheme presended in \cite {b9}. The relevant wave function obeys eq. (3.3) of \cite {b9} with a=2 and is 
given by eq. (3.8) therein
\begin{eqnarray}
&& \psi (x)=e^{-ipx}F(1,-1,1-\frac{E_p}{m_2} \; ;y) \label{w32} \\ 
&& y\equiv -\frac{\lambda }{48m_2^2}\frac{z_2^2}{1-\frac{\lambda }{48m_2^2}z_2^2} \label{w33}
\end{eqnarray}
or, explicitly
\begin{eqnarray}
\psi (x)=e^{-ipx}\left(\frac{1-\frac{\lambda }{48m_2^2}(\frac{E_p+m_2}{E_P-m_2})z_2^2}{1-\frac{\lambda }
{48m_2^2}z_2^2}\right) \label{w34}
\end{eqnarray}
Eqs. (\ref{w31}) and (\ref{w34}) can be now compared. According to the general reasoning presented in Sec.II one should 
identify $ E_p\leftrightarrow m_1$\ and neglect space-dependent factor in (\ref{w34}). Then we see that (\ref{w31}) and 
(\ref{w34}) indeed coincide (one should also take into account the difference in wave function normalization).

In particular, the nuliffication of $ 2\rightarrow n$\ hard-soft amplitudes results from Ward identities for purely 
soft ones \cite  {b2}.
\subsection{Henon-Heiles system}
Consider the theory given by the lagrangian
\begin{eqnarray}
&& L=\frac{1}{2}(\partial _{\mu }\Phi _1\partial ^{\mu }\Phi _1-m_1^2\Phi _1^2)+\frac{1}{2}(\partial _{\mu }\Phi _2\partial ^
{\mu }\Phi _2-m_2^2\Phi _2^2)+ \nonumber \\
&& -\frac{g}{2!}\Phi _1^2\Phi _2-\frac{\lambda }{3!}\Phi _2^3 \label{w35}
\end{eqnarray}
It is symetric under $ \Phi _1\rightarrow -\Phi _1, \;\;  \Phi _2\rightarrow \Phi _2$. There exists no stable ground 
state here but on the perturbative level theory is perfectly well defined. We shall analyse its threshold behaviour 
in few steps. \\
{\large \bf (A) $ 2\rightarrow n$\ amplitudes: $ \Phi _1$\ -particles}

We assume that both initial particles are carried by $ \Phi _1$\ field. According to the procedure outlined 
in Sec.II we have to solve first the field equation for $ \Phi _2$\ assuming $ \Phi _1\equiv 0$\ and no space-dependence,
\begin{eqnarray}
&& (\partial _t^2+m_2^2)\Phi _2+\frac{\lambda }{2}\Phi _2^2=0 \label{w36} \\
&& \Phi _2\mid _{\lambda =0}=ze^{im_2t}\equiv \frac{{\overline \beta }}{\sqrt{(2\pi )^32m_2}}e^{im_2t} \label{w37}
\end{eqnarray}
The proper solution ( i.e. the one reproducing tree-graph expansion ) is obtained by assuming the total energy to 
vanish; it reads
\begin{eqnarray}
\Phi _2(t)=\frac{ze^{im_2t}}{(1-\frac{z\lambda }{12m_2^2}e^{im_2t})^2} \label{w38}
\end{eqnarray}
This is the generating function for $ \Phi ^3$\ -theory. The counterpart of eq. (\ref{w23}) takes the form
\begin{eqnarray}
(\Box +m_1^2+g\Phi _2(t))\psi (x)=0 \label{w39}
\end{eqnarray}
Puting
\begin{eqnarray}
\psi (x)=\frac{1}{\sqrt{(2\pi )^32E_p}}e^{-ipx}\tilde \psi (t), \;\; p^2=m_1^2, \;\; E_p\equiv \sqrt{\vec p^2+m_1^2} 
\label{w40}
\end{eqnarray}
we arrive at
\begin{eqnarray}
&& (\partial _t^2-2iE_p\partial _t+g\Phi _2(t))\tilde \psi (t)=0 \label{w41} \\
 && \tilde \psi (t)\mid _{g=0}=1 \label{w42}
\end{eqnarray}
In order to solve (\ref{w41}) we make a change of variable \cite {b9}
\begin{eqnarray}
y=\frac{\frac{-z\lambda }{12m_2^2}e^{im_2t}}{1-\frac{z\lambda }{12m_2^2}e^{im_2t}} \label{w43}
\end{eqnarray}
which converts (\ref{w41}) into 
\begin{eqnarray}
\left(y(1-y)\frac{d^2}{dy^2}+((1-\frac{2E_p}{m_2})-2y)\frac{d}{dy}+\frac{12g}{\lambda }\right)\tilde \psi (y)=0 \label{w44}
\end{eqnarray}
The relevant solution reads
\begin{eqnarray}
\tilde \psi (t)=F(a_1,a_2, \;\;1-\frac{2E_p}{m_2};y), \label{w45}
\end{eqnarray}
where $F$\ is the hypergeometric function and 
\begin{eqnarray}
a_{1,2}=\frac{1\pm \sqrt{1+\frac{48g}{\lambda }}}{2} \label{w46}
\end{eqnarray}
Let us now apply the LSZ-reduction to produce the second initial particle:
\begin{eqnarray}
&& \frac{i}{\sqrt{(2\pi )^32E_q}}\int d^4xe^{-iqx}(\Box +m_1^2)\psi (x)= \nonumber \\
&& =\frac{i\delta ^{(3)}(\vec p+\vec q)}{2E_p}\int\limits_{-\infty}^{\infty}dte^{-iE_pt}(\partial _t^2+E_p^2)
(e^{-iE_pt}\tilde \psi (t)) \label{w47}
\end{eqnarray}
Now, from (\ref{w47}) we see that $ \tilde \psi (t)$\ has the following exspansion 
\begin{eqnarray}
\tilde \psi (t)=\sum\limits_{n\geq 0}\alpha _ne^{inm_2t}  \label{w48}
\end{eqnarray}
Taking into account the contribution from n-th term one gets
\begin{eqnarray}
&& \frac{i}{\sqrt{(2\pi )^32E_q}}\int d^4xe^{-iqx}(\Box  +m_1^2)\psi (x)\mid _n= \nonumber \\
&& =2i\pi \delta ^{(3)}(\vec p+\vec q)\delta (2E_p-nm_2)(2E_p-nm_2)\alpha_n \;; \label{w49}
\end{eqnarray}
the conribution is nonvanishing only provided $\alpha  _n$\ has a pole at $ 2E_p=nm_2$. This pole term is easily identified 
from the solution (\ref{w45}). The result reads
\begin{eqnarray}
A(2\rightarrow n)=\frac{4i\pi \delta ^{(3)}(\vec p+\vec q)\delta (2E_p-nm_2)E_p}{((2\pi )^32m_2)^{\frac{n}{2}}n!}
(\frac{\lambda }{12m_2^2})^n\frac{\Gamma (a_1+n)\Gamma (a_2+n)}{\Gamma (a_1)\Gamma (a_2)} \label{w50}
\end{eqnarray}
Eq.(\ref{w50}) can be checked by Feynman-graph method. The relevant Feynman rules read
\begin{fmffile}{fgraph}
\begin{equation*}
\parbox{28mm}{
\begin{fmfgraph*}(65,20)
\fmfforce{0w,.5h}{v}
\fmfforce{1w,.5h}{x}
\fmf{plain}{v,x}
\fmfv{label=$1$}{v}
\fmfv{label=$1$}{x}
\end{fmfgraph*}
}
\;\;\;\;\frac{i}{p^2-m_1^2+i\varepsilon} 
\end{equation*}
\begin{equation*}
\parbox{28mm}{
\begin{fmfgraph*}(65,20)
\fmfforce{0w,.5h}{v}
\fmfforce{1w,.5h}{x}
\fmf{dashes}{v,x}
\fmfv{label=$2$}{v}
\fmfv{label=$2$}{x}
\end{fmfgraph*}
}
\;\;\;\;\frac{i}{p^2-m_2^2+i\varepsilon} 
\end{equation*}
\\
\begin{equation*}
\parbox{45mm}{
\begin{fmfgraph*}(65,30)
\fmfforce{0w,1.2h}{x}
\fmfforce{1w,1.2h}{y}
\fmfforce{.5w,.6h}{z}
\fmfforce{.5w,-.3h}{t}
\fmf{plain}{x,z}
\fmf{plain}{z,y}
\fmf{dashes}{z,t}
\fmfv{label=$1$}{x}
\fmfv{label=$1$}{y}
\fmfv{label=$2$}{t}
\end{fmfgraph*}
}
-ig 
\end{equation*}
\\
\\
\begin{equation*}
\parbox{45mm}{
\begin{fmfgraph*}(65,30)
\fmfforce{0w,1.2h}{x}
\fmfforce{1w,1.2h}{y}
\fmfforce{.5w,.6h}{z}
\fmfforce{.5w,-.3h}{t}
\fmf{dashes}{x,z}
\fmf{dashes}{z,y}
\fmf{dashes}{z,t}
\fmfv{label=$2$}{x}
\fmfv{label=$2$}{y}
\fmfv{label=$2$}{t}
\end{fmfgraph*}
}
-i\lambda  
\end{equation*}
\\

For $n=1$\ there is only one contribution coming from the elementary vertex
\\
\\
\begin{equation*}
\parbox{35mm}{
\begin{fmfgraph*}(65,30)
\fmfforce{0w,1h}{x}
\fmfforce{0w,0h}{y}
\fmfforce{.5w,.5h}{z}
\fmfforce{1.2w,.5h}{t}
\fmf{plain}{x,z}
\fmf{plain}{z,y}
\fmf{dashes}{z,t}
\fmfv{label=$1$}{x}
\fmfv{label=$1$}{y}
\fmfv{label=$2$}{t}
\end{fmfgraph*}
}
\end{equation*}
\\
which contributes an amount $\frac{-ig\pi \delta (2E_p-m_2)\delta^{(3)}(\vec{p}+\vec{q})}{\sqrt{(2\pi )^32m_2}E_p}$,
in agreement with (\ref{w50}) for $n=1$.

The case $n=2$\ is slightly more complicated . There are three graphs contributing
\begin{equation*}
\parbox{35mm}{
\begin{fmfgraph*}(65,30)
\fmfforce{0w,1h}{x}
\fmfforce{0w,0h}{y}
\fmfforce{.4w,.8h}{z}
\fmfforce{.4w,.2h}{t}
\fmfforce{1w,.8h}{a}
\fmfforce{1w,.2h}{b}
\fmf{plain}{x,z}
\fmf{plain}{t,y}
\fmf{plain}{z,t}
\fmf{dashes}{z,a}
\fmf{dashes}{t,b}
\fmfv{label=$1$}{x}
\fmfv{label=$1$}{y}
\fmfv{label=$2$}{a}
\fmfv{label=$2$}{b}
\end{fmfgraph*}
}+\;\;\;
\parbox{35mm}{
\begin{fmfgraph*}(65,30)
\fmfforce{0w,1h}{x}
\fmfforce{0w,0h}{y}
\fmfforce{.4w,.8h}{z}
\fmfforce{.4w,.2h}{t}
\fmfforce{1w,.8h}{a}
\fmfforce{1w,.2h}{b}
\fmf{plain}{x,z}
\fmf{plain}{t,y}
\fmf{plain}{z,t}
\fmf{dashes}{z,b}
\fmf{dashes}{t,a}
\fmfv{label=$1$}{x}
\fmfv{label=$1$}{y}
\fmfv{label=$2$}{a}
\fmfv{label=$2$}{b}
\end{fmfgraph*}
}+\;\;\;
\parbox{35mm}{
\begin{fmfgraph*}(65,30)
\fmfforce{0w,1h}{x}
\fmfforce{0w,0h}{y}
\fmfforce{.4w,.5h}{z}
\fmfforce{.8w,.5h}{t}
\fmfforce{1w,1h}{a}
\fmfforce{1w,0h}{b}
\fmf{plain}{x,z}
\fmf{plain}{z,y}
\fmf{dashes}{z,t}
\fmf{dashes}{a,t}
\fmf{dashes}{b,t}
\fmfv{label=$1$}{x}
\fmfv{label=$1$}{y}
\fmfv{label=$2$}{a}
\fmfv{label=$2$}{b}
\end{fmfgraph*}
}
\end{equation*}
\\
and they sum up to
\begin{eqnarray}
\frac{i\delta^{(3)}(\vec{p}+\vec{q})\delta (2E_p-2m_2)g}{(2\pi)^22m_2^4}\left(g-\frac{\lambda}{6}\right)\label{w51}
\end{eqnarray}
which again agrees with general formula (\ref{w50}). 

Let us now note the following: for
\begin{eqnarray}
\frac{g}{\lambda}=\frac{N(N+1)}{12}\label{w52}
\end{eqnarray}
one finds $a_2=-N$. Therefore, due to $\Gamma (a_2+n)/\Gamma (a_2)=(-N)(1-N)...(n-1-N)$, all amplitudes 
$A(2\rightarrow n),\;n\geq N+1$, vanish. In particular, for $N=1$\ the only nonvanishing amplitude corresponds to $n=1$. 
This latter result can be understood by invoking tyhe equivalence with purely threshold amplitudes, as discussed in
Sec. II. In fact, $N=1$\ implies, through (\ref{w52}), $g=\frac{\lambda}{6}$; in this case the Henon-Heiles
system becomes integrable and separable in parabolic coordinates \cite{b11}. Let us consider the relevant solution
in more detail.
\\
{\large \bf (B) Henon-Heiles model: integrable case}

If $g=\frac{\lambda}{6}$\ the reduced dynamics becomes separable in parabolic coordinates
\begin{eqnarray}
\Phi_1^2=-4\zeta \eta\;\;\;\;\;\; \nonumber \\
&&\kappa \equiv \frac{3}{\lambda}(4m_1^2-m_2^2) \label{w53} \\
\Phi_2=\zeta +\eta +\kappa \nonumber 
\end{eqnarray}
The additional integral of motion resulting from the separation of variables reads
\begin{eqnarray}
&&F=\frac{\lambda}{2}\dot{\Phi}_1(\Phi_1\dot{\Phi}_2-\Phi_2\dot{\Phi_1})+\frac{\lambda \kappa}{2}(\dot{\Phi}_1^2+m_1^2\Phi_1^2)
+\frac{\lambda m_1^2}{2}\Phi_1^2\Phi_2+ \nonumber \\
&&+\frac{\lambda^2}{24}\Phi_1^2\left(\frac{\Phi_1^2}{4}+\Phi_2^2\right) \label{w54}
\end{eqnarray}
$F$\ generates the following symmetry transformation
\begin{eqnarray}
&&\Phi_1\rightarrow \Phi_1'=\Phi_1+\varepsilon \lambda (\kappa \dot{\Phi}_1+\frac{1}{2}
(\Phi_1\dot{\Phi}_2-2\Phi_2\dot{\Phi}_1))\label{w55} \\
&&\Phi_2\rightarrow \Phi_2'=\Phi_2+\frac{\varepsilon \lambda}{2}\Phi_1\dot{\Phi}_1 \nonumber
\end{eqnarray}
We are looking for the solution obeying 
\begin{eqnarray}
\Phi_1\mid_{\lambda =0}=z_1e^{-im_1t} \nonumber \\
&&z_i\equiv \frac{\beta_i}{\sqrt{(2\pi )^32m_i}}\label{w56} \\
\Phi_2\mid_{\lambda =0}=z_2e^{im_2t}\; \nonumber
\end{eqnarray}
It corresponds to $E=0$\ and $F=0$\ and reads
\begin{eqnarray}
&&\Phi_1=\frac{2m_1(4m_1^2-m_2^2)(1-x)y}{\lambda (2m_1(1-x)(1-y^2)+m_2(1+x)(1+y^2))}\label{w57} \\
&&\Phi_2=\frac{4\kappa (m_2^2x(1+y^2)^2+4m_1^2y^2(1-x)^2)}{(2m_1(1-x)(1-y^2)+m_2(1+x)(1+y^2))^2}\label{w58}
\end{eqnarray}
where
\begin{eqnarray}
&&x=\frac{z_2\lambda}{m_2^2}\left(\frac{2m_1+m_2}{2m_1-m_2}\right)e^{im_2t}\label{w59} \\
&&y=\frac{z_1\lambda e^{-im_1t}}{2m_1(2m_1-m_2)}\label{w60}
\end{eqnarray}
Let us analyse this solution in some detail. For $2m_1\neq m_2$\ there are no resonances so the relevant threshold
amplitudes vanish. On the other hand,
\begin{eqnarray}
\frac{\partial \Phi_1}{\partial z_1}\mid_{z_1=0}=\frac{1-\left(\frac{2m_1+m_2}{2m_1-m_2}\right)
\frac{z_2\lambda}{m_2^2}e^{im_2t}}{1-\frac{z_2\lambda}{m_2^2}e^{im_2t}}\simeq 1-\frac{2\lambda z_2e^{im_2t}}{m_2(
2m_1-m_2)}+O(z_2^2)\label{w61}
\end{eqnarray}
which gives nonzero amplitude $2\rightarrow 1$\ for $2m_1=m_2$. This is in agreement with general arguments concerning
the relation between symmetries and amplitude nullification \cite{b10}, \cite{b2}; in fact, for $2m_1=m_2$\ the relevant
linear part in the transformation formulae (\ref{w55}) is absent.

By replacing $m_1$\ by $E_p\equiv \sqrt{\vec{p}^2+m_1^2}$ our expression (\ref{w61}) coincides with (\ref{w45}) calculated 
for $N=1$. Therefore, the symmetry (\ref{w55}), which is valid for arbitrary masses, inplies the vanishing of 
$2\rightarrow n$\ hard- soft amplitudes. 
\newpage
\section{Fermions}
$ $\

It is straightforward to include higher-spin bosons into our scheme. As far as fermions are concerned the only but 
crucial modification is to replace the parameters $\beta,\;\bar{\beta}$\ entering (\ref{w2}) by anticommuting Grassman
variables; this allows to implement Pauli exlusion principle.

As an example we shall compute the threshold amplitudes for $n$-boson production by fermion-antifermion pair. The theory we
are considering provides a toy model for fermionic mass generation is the standard model via spontaneous symmetry breaking.
It has been considered from the point of view of amplitudes nullification in Refs. \cite{b4}, \cite{b5}; diagrammatic
approach through recurrence relations has been developed in \cite{b7}.

Our model contains massless fermionic field coupled by Yukawa term to the Higgs field. The relevant lagrangian reads
\begin{eqnarray}
L=i\bar{\psi}\gamma^{\mu}\partial_{\mu}\psi+\frac{1}{2}(\partial_{\mu}\Phi \partial^{\mu}\Phi +m^2\Phi^2)-
\frac{\lambda}{4!}\Phi^4-g\Phi \psi \bar{\psi}\label{w62}
\end{eqnarray}

Due to the wrong sign of mass term the vacuum solution is nontrivial
\begin{eqnarray}
\psi =0,\;\bar{\psi}=0,\;\Phi=v\equiv \sqrt{\frac{3!m^2}{\lambda}}\label{w63}
\end{eqnarray}
Define the physical field $\rho$\ by
\begin{eqnarray}
\rho \equiv \Phi -v \label{w64}
\end{eqnarray}
In terms of this new field the lagrangian reads
\begin{eqnarray}
&&L=\bar{\psi}(i\gamma^{\mu}\partial_{\mu}-M)\psi+\frac{1}{2}(\partial_{\mu}\rho \partial^{\mu}\rho -m_{\rho}^2
\rho^2)-\frac{\lambda v}{3!}\rho^3-\frac{\lambda}{4!}\rho^4+ \nonumber \\
&&-g\rho \bar{\psi} \psi ;\label{w65}
\end{eqnarray}
here $M=gv,\;m_{\rho}=\sqrt{2}m$\ are fermionic and bosonic masses, respectively.

$L$\ is invariant under $\psi \rightarrow -\psi,\;\bar{\psi}\rightarrow -\bar{\psi},\;\rho \rightarrow \rho$; therefore,
one can apply the strategy of Sec. II. First, we solve field equations for $\rho$\ under the condition 
$\psi =\bar{\psi}=0$\ and
\begin{eqnarray}
\rho \mid_{\lambda =0}=\frac{\beta e^{im_{\rho}t}}{\sqrt{(2\pi )^32m_{\rho}}}\equiv z_0e^{im_{\rho}t}\equiv z(t) \label{w66}
\end{eqnarray}
The solution reads \cite{b9}
\begin{eqnarray}
\rho (t)=\frac{z(t)}{1-\frac{z(t)}{2v}} \label{w67}
\end{eqnarray}
or
\begin{eqnarray}
\Phi (t)=v\frac{1+\frac{z(t)}{2v}}{1-\frac{z(t)}{2v}}\label{w68}
\end{eqnarray}
The counterpart of (\ref{w23}) takes now the form
\begin{eqnarray}
&&(i\gamma^{\mu}\partial_{\mu}-g\Phi (t))\psi_{ps}(x)=0\label{w69}\\
&&\psi_{ps}(x)\mid_{{g=0\atop gv=M}}=\sqrt{\frac{M}{(2\pi )^3E_p}}u(p,s)e^{-ipx}\label{w70}
\end{eqnarray}
The same reasoning as in bosonic case suggests the following Ansatz for $\psi_{ps}$:
\begin{eqnarray}
\psi_{ps}(x)={\zeta F(t)\choose \eta \tilde{F}(t)}e^{ip^kx^k}e^{-iE_pt},\label{w71}
\end{eqnarray}
where 
\begin{eqnarray}
{\zeta \choose \eta}=\sqrt{\frac{M}{(2\pi )^3E_p}}u(p,s) \label{w72}
\end{eqnarray}
is the standard solution to Dirac equation \cite{b13}. Inserting the Ansatz (\ref{w71}), (\ref{w72}) into 
eq. (\ref{w69}) we arrive at
\begin{eqnarray}
&&(E_p+i\partial_t-g\Phi (t))F(t)\zeta +\tilde{F}(t)p_k\sigma^k\eta =0\label{w73}\\
&&(-E_p-i\partial_t-g\Phi (t))\tilde{F}(t)\eta -F(t)p_k\sigma^k\zeta =0\label{w74}
\end{eqnarray}
or, using (\ref{w72}),
\begin{eqnarray}
&&(E_p+i\partial_t-g\Phi (t))F(t)-(E-M)\tilde{F}(t)=0\label{w75} \\
&&(E_p+i\partial_t+g\Phi (t))\tilde{F}(t)-(E+M)F(t)=0;\label{w76} 
\end{eqnarray}
the boundary condition (\ref{w70}) is obeyed provided 
\begin{eqnarray}
F\mid_{{g=0\atop gv=M}}=1,\;\tilde{F}\mid_{{g=0\atop gv=M}}=1\label{w77}
\end{eqnarray}

It is not difficult to find the solution to (\ref{w75}), (\ref{w76}) obeying (\ref{w77}). We apply 
$(E_p+i\partial_t+g\Phi (t))$\ to (\ref{w75}) and use (\ref{w76}); then 
\begin{eqnarray}
(\partial_t^2-2iE_p\partial_t+g^2\Phi^2(t)+ig\partial_t\Phi (t)+M^2)F=0\label{w78}
\end{eqnarray}
Changing the variable,
\begin{eqnarray}
y=\frac{-\frac{z(t)}{2v}}{1-\frac{z(t)}{2v}},\label{w79}
\end{eqnarray}
we arrive at the hipergeometric equation. Its solution reads
\begin{eqnarray}
&&F(t)=F(\alpha ,\;\beta ,\;\gamma ;\;y)\label{w80}\\
&&\alpha \equiv \frac{2M}{m_{\rho}},\;\beta \equiv 1-\frac{2M}{m_{\rho}},\;\gamma \equiv 1-\frac{2E_p}{m_{\rho}}\label{w81}
\end{eqnarray}
For $\tilde{F}(t)$\ we obtain, respectively,
\begin{eqnarray}
\tilde{F}(t)=F(\alpha +1,\;\beta -1,\;\gamma ;\;y);\label{w82}
\end{eqnarray}
(\ref{w80}) and (\ref{w82}) satisfy (\ref{w75}), (\ref{w76}).

We are now ready to calculate the exact expression for the tree-graph threshold amplitude $A(f\bar{f}\rightarrow n\rho )$.
To this end we reduce the fermionic antiparticle by LSZ formula. Then
\begin{eqnarray}
-i\int d^4x\bar{v}_{p's'}(x)(i\gamma^{\mu}\partial_{\mu}-M)\psi_{ps}(x)\label{w83}
\end{eqnarray}
becomes the generating functional for all such amplitudes; here
\begin{eqnarray}
\bar{v}_{p's'}(x)=\sqrt{\frac{M}{(2\pi )^3E_{p'}}}\bar{v}(p',s')e^{-ip'x}\label{w84}
\end{eqnarray}
In order to calculate (\ref{w83}) let us write
\begin{eqnarray}
\psi_{ps}(x)=\sqrt{\frac{M}{(2\pi )^3E_p}}\tilde{\psi}(x^0)e^{-ipx};\label{w85}
\end{eqnarray}
then (\ref{w83}) can be rewritten as follows
\begin{eqnarray}
&&\frac{-iM}{(2\pi )^3\sqrt{E_p E_{p'}}}\int d^4x\bar{v}(p',s')e^{-ip'x}(i\gamma^{\mu}\partial_{\mu}-M)
\tilde{\psi}(x^0)e^{-ipx}=\nonumber \\
&&\frac{-iM}{(2\pi )^3\sqrt{E_{p'} E_{p'}}}\int d^4x\bar{v}(p',s')e^{-i(p+p')x}(i\gamma^{\mu}p_{\mu}-M+i\gamma^0\partial_0)
\tilde{\psi}(x^0)=\nonumber \\
&&\frac{-iM}{E_p}\delta^{(3)}(\vec{p}+\vec{p}')\int\limits_{-\infty}^{\infty}dx^0e^{-2iE_px^0}\bar{v}(p',s')
(\gamma^{\mu}p_{\mu}-M+i\gamma^0\partial_0 )\tilde{\psi}(x^0)\label{w86}
\end{eqnarray}
Let us note that, due to $\vec{p}\;'=-\vec{p}\;'$,
\begin{eqnarray}
\bar{v}(p',s')(\gamma^{\mu}p_{\mu}-M)=2E_p\bar{v}(p',s')\gamma^0 \label{w87}
\end{eqnarray}
and our functional takes the form
\begin{eqnarray}
\frac{-iM}{E_p}\delta^{(3)}(\vec{p}+\vec{p}')\int\limits_{-\infty}^{\infty}dx^0e^{-2iE_px^0}\bar{v}(p',s')\gamma^0
(2E_p+i\partial_0)\tilde{\psi}(x^0)\label{w88}
\end{eqnarray} 
Consider the last expression. It can be analysed in a similar way as in  purely bosonic case. In order to isolate
the amplitude for the creation of $n$\ bosons one has to single out the term proportional to $z_0^n$. Due to the structure
of the solution $\psi_{ps}(x)$\ it can be expanded in Fourier series in $e^{im_{\rho}t}$. The term $e^{inm_{\rho}t}$\ 
produces the energy delta function $\delta (2E_p-nm_{\rho})$; however, because of time derivative in (\ref{w88}) this delta 
function is accompanied by $2E_p-nm_{\rho}$, so the result is zero unless there is an additional pole coming from the gamma 
functions entering hipergeometric series. It is straightforward to isolate this term arriving at the following final
expression for the amplitude
\begin{eqnarray}
&&A(f\bar{f}\rightarrow n\rho )=\label{w89} \\
&&\frac{-4\pi iM\delta (2E_p-nm_{\rho})\delta^{(3)}(\vec{p}+\vec{p}')}{(
\sqrt{(2\pi )^32m_{\rho})}^n(n-1)!(2v)^n}\frac{\Gamma (n+\frac{2M}{m_{\rho}})\Gamma (n-\frac{2M}{m_{\rho}})}
{\Gamma (\frac{2M}{m_{\rho}})\Gamma (1-\frac{2M}{m_{\rho}})}
\bar{v}(p',s')u(p,s)\nonumber
\end{eqnarray}
The spin structure of this formula is dictated by invariance properties, angular momentum and parity conservation. In order 
to check the formula (\ref{w89}) we compare it with Feynman-graph computation. For $n=2$\ eq. (\ref{w89}) takes the form
\begin{eqnarray}
&&A(f\bar{f}\rightarrow 2\rho )= \nonumber \\
&&\frac{-i}{4\pi^2}\delta (2E_p-2m_{\rho})\delta^{(3)}(\vec{p}+\vec{p}')
\left(\frac{M^2}{2m^2_{\rho}v^2}-\frac{2M^4}{m_{\rho}^4v^2}\right)\bar{v}(p',s')u(p,s)\label{w90}
\end{eqnarray}
The relevant Feynman rules are
\begin{equation*}
\parbox{28mm}{
\begin{fmfgraph*}(65,20)
\fmfforce{0w,.5h}{v}
\fmfforce{1w,.5h}{x}
\fmf{plain_arrow}{v,x}
\end{fmfgraph*}
}
\;\;\;\;\frac{i(\gamma^{\mu}p_{\mu} +M)}{p^2-M^2+i\varepsilon} 
\end{equation*}
\begin{equation*}
\parbox{28mm}{
\begin{fmfgraph*}(65,20)
\fmfforce{0w,.5h}{v}
\fmfforce{1w,.5h}{x}
\fmf{dashes}{v,x}
\end{fmfgraph*}
}
\;\;\;\;\frac{i}{p^2-m_{\rho}^2+i\varepsilon} 
\end{equation*}
\\
\begin{equation*}
\parbox{28mm}{
\begin{fmfgraph*}(65,45)
\fmfforce{0w,1h}{x}
\fmfforce{0w,0h}{y}
\fmfforce{.5w,.5h}{z}
\fmfforce{1.2w,.5h}{t}
\fmf{plain_arrow}{z,x}
\fmf{plain_arrow}{y,z}
\fmf{dashes}{z,t}
\end{fmfgraph*}
}-ig
\end{equation*}
\begin{equation*}
\parbox{28mm}{
\begin{fmfgraph*}(65,45)
\fmfforce{0w,1h}{x}
\fmfforce{0w,0h}{y}
\fmfforce{.5w,.5h}{z}
\fmfforce{1.2w,.5h}{t}
\fmf{dashes}{x,z}
\fmf{dashes}{z,y}
\fmf{dashes}{z,t}
\end{fmfgraph*}
}-i\lambda v
\end{equation*}\\
\begin{equation*}
\parbox{28mm}{
\begin{fmfgraph*}(65,45)
\fmfforce{0w,1h}{x}
\fmfforce{0w,0h}{y}
\fmfforce{1w,0h}{z}
\fmfforce{1w,1h}{t}
\fmf{dashes}{x,z}
\fmf{dashes}{t,y}
\end{fmfgraph*}
}-i\lambda
\end{equation*}
\begin{equation*}
Fig. 2
\end{equation*}
The graphs contributing to $A(f\bar{f}\rightarrow 2\rho )$\ are depicted on Fig. 3
\begin{equation*}
\parbox{35mm}{
\begin{fmfgraph*}(65,40)
\fmfforce{0w,1h}{x}
\fmfforce{0w,0h}{y}
\fmfforce{.4w,.8h}{z}
\fmfforce{.4w,.2h}{t}
\fmfforce{1w,.8h}{a}
\fmfforce{1w,.2h}{b}
\fmf{plain_arrow}{z,x}
\fmf{plain_arrow}{y,t}
\fmf{plain_arrow}{t,z}
\fmf{dashes}{z,a}
\fmf{dashes}{t,b}
\end{fmfgraph*}
}+\;\;\;
\parbox{35mm}{
\begin{fmfgraph*}(65,40)
\fmfforce{0w,1h}{x}
\fmfforce{0w,0h}{y}
\fmfforce{.4w,.8h}{z}
\fmfforce{.4w,.2h}{t}
\fmfforce{1w,.8h}{a}
\fmfforce{1w,.2h}{b}
\fmf{plain_arrow}{z,x}
\fmf{plain_arrow}{y,t}
\fmf{plain_arrow}{t,z}
\fmf{dashes}{z,b}
\fmf{dashes}{t,a}
\end{fmfgraph*}
}+\;\;\;
\parbox{35mm}{
\begin{fmfgraph*}(65,40)
\fmfforce{0w,1h}{x}
\fmfforce{0w,0h}{y}
\fmfforce{.4w,.5h}{z}
\fmfforce{.8w,.5h}{t}
\fmfforce{1w,1h}{a}
\fmfforce{1w,0h}{b}
\fmf{plain_arrow}{z,x}
\fmf{plain_arrow}{y,z}
\fmf{dashes}{z,t}
\fmf{dashes}{a,t}
\fmf{dashes}{b,t}
\end{fmfgraph*}
}
\end{equation*}
\begin{equation*}
Fig. 3
\end{equation*}

Again, the amplitude exhibits nullification phenomenon \cite{b4}, \cite{b5}, \cite{b7}, \cite{b9}. Due to 
\begin{eqnarray}
\frac{\Gamma (n-\alpha )}{\Gamma (1-\alpha )}=(1-\alpha )(2-\alpha )...(n-1-\alpha )\label{w91}
\end{eqnarray}
the amplitude vanishes for $n>N$\ provided $\alpha \equiv \frac{2M}{m_{\rho}}=N$\ is an integer (actually, $n\leq N$\
amplitudes also vanish due to energy conservation and the properties of spinor wave functions). In particular, for $N=1$\
all amplitudes, vanish.
\section{Final remarks}
$ $\

We presented a simple derivation of the Brown-Zhai Feynman wave function algorithm for general theories. It allows to
compare our process with a purely soft one. The important point is that, for the latter, we can use Ward identities
following from the integrability of reduced dynamics \cite{b2} to prove their vanishing. This argument works in purely 
scalar case (cf. also \cite{b2}) as well as fermionic one \cite{b14}. In this way we identify, in some cases, the symmetry
underlying nullification of $2\rightarrow n$\ amplitudes.
\newpage
\end{fmffile}

\end{document}